\documentclass[12pt,preprint]{aastex}

\begin{document}

\title{Stellar Abundances and Molecular Hydrogen in High-Redshift Galaxies --
the Far-Ultraviolet View 
\footnote{Based on observations made with the NASA-CNES-CSA Far Ultraviolet 
Spectroscopic Explorer. FUSE is operated for NASA by the Johns Hopkins 
University under NASA contract NAS5-32985.}}

\author{William C. Keel}
\affil{Department of Physics and Astronomy, University of Alabama, Box 870324,
Tuscaloosa, AL 35487; keel@bildad.astr.ua.edu}

\begin{abstract}
FUSE spectra of star-forming regions in nearby galaxies are compared to
composite spectra of Lyman-break galaxies (LBGs), binned by strength of
Lyman $\alpha$ emission and by mid-UV luminosity. Several far-UV spectral
features, including lines dominated by stellar wind and 
by photospheric components,
are very sensitive to stellar abundances. Their measurement in Lyman-break
galaxies is compromised by the strong interstellar absorption features,
allowing in some cases only upper limits.
The derived
C and N abundances in the LBGs are no higher than half solar (scaled to
oxygen abundance for comparison with emission-line analyses), independent
of the strength of Lyman $\alpha$ emission. P V absorption indicates
abundances as low as 0.1 solar, with an upper limit near
0.4 solar in the reddest and weakest-emission galaxies. Unresolved
interstellar absorption components would further lower the derived
abundances.
Trends of line strength, and derived abundances, are stronger with 
mid-UV luminosity than with Lyman-$\alpha$ strength. 
H$_2$ absorption in the Lyman and Werner bands is very weak in the
LBGs. Template H$_2$ absorption spectra convolved to appropriate
resolution show that strict upper limits N(H$_2$) $ < 10^{18}$ cm$^{-2}$
apply in all cases, with more stringent values 
appropriate for the stronger-emission composites and for mixes of
H$_2$ level populations 
like those on Milky Way sight lines. Since the
UV-bright regions are likely to be widespread in these galaxies,
these results rule out massive diffuse reservoirs of H$_2$, and suggest that the
dust/gas ratio is already fairly large at $z \approx 3$.
\end{abstract}


\section{Introduction}

Massive stars play crucial roles in galaxy evolution, quite aside from
dominating many of the observed properties of star-forming systems.
Their mass loss and energy input are major factors in the history of
the interstellar medium. The far-ultraviolet range, between the
Lyman limit and Lyman $\alpha$, shows us only these stars, giving us
a direct window to the ongoing star formation wherever the extinction is
low enough. The far-UV range provides an embarrassment of riches for
tracing stellar winds and the interstellar medium in both atomic and
molecular form, including atomic species and ionization stages not
represented at longer wavelengths. Both extinction 
and the history of star formation make only subtle
differences in the observed far-UV spectrum, in each case because the parameter 
range over which stars contribute to the far-UV
light is rather limited.

The rich information content of FUSE spectra has revolutionized our
understanding of the UV universe, particularly including star-forming systems.
In this paper, I compare properties of local star-forming galaxies, whose
chemical compositions are well established from emission-line spectroscopy,
to those of Lyman-break galaxies at $z \approx 3$. This comparison
allows estimates of the metallicity in the massive stars, and
sets interesting limits on absorption from surrounding molecular gas. 

\section{Data}

\subsection{FUSE spectra}

The local reference data are FUSE observations of luminous star-forming
regions in nearby galaxies, described by Keel, Holberg, \& Treuthardt (2004).
These consist of the four UV-brightest giant H II regions in M33 (NGC 588, 592,
595, and 604), a bright outer H II region in M101 (NGC 5461), the
starburst nucleus of NGC 7714, and the summed spectrum of I Zw 18 
produced by Aloisi et al. (2003). These spectra were all obtained with the
large 30" aperture, encompassing essentially the entire young populations
with the goal of comparison with integrated spectra of distant galaxies.
Exposure times for the new observations were 4000-7000 seconds, while
I Zw 18 was observed for a total of 95,000 seconds. Since NGC 604 is
an order of magnitude brighter than the other objects in the FUV band,
its spectral signal-to-noise ratio is the best of the whole sample, and
the data can be evaluated at higher resolution (0.1 \AA\ ). As detailed
by Keel et al., Galactic H$_2$ lines were removed by fitting families of
Gaussian or Voigt profiles (for the stronger transitions) to the various
H$_2$ multiplets, with line identifications from the H$_2$OOLS package
by McCandliss (2003). Several atomic ISM lines from the Milky Way and
high-velocity clouds were similarly excised. 

Fig. 1 identifies
important interstellar and wind features in the far-UV spectral
range on the FUSE spectra of NGC 604 and I Zw 18. These objects
play special roles in our understanding of hot stellar populations
in the far-UV (and indeed at other wavelengths as well). NGC 604 is
by far the most luminous star-forming region in the nearby spiral
NGC 598 (M33), comparably powerful to the 30 Doradus complex and 
perhaps NGC 3603 in our own Galaxy. However, it does not contain
a central super star cluster, but is instead a collection of smaller
and less dense subsystems, whose star-forming history featured
a strong burst about 3 Myr ago (Gonz{\'a}lez Delgado \& P{\'e}rez 2000)).
Active star formation has proceeded long enough in NGC 604 to
generate an overall bubble geometry in the surrounding H II region, one
so extensive that it probably contributes to the low limits on
foreground H$_2$ through association or displacement of the
surrounding disk ISM (Bosch et al. 2002). The optical emission
lines yield an oxygen abundance O/H=0.66 solar (Vilchez et al. 1988).
In contrast to this relatively metal-rich object, I Zw 18 remains the
galaxy with the lowest known emission-line abundances (O/H only 0.05 solar), 
more than 30 years after
the measurements of Searle \& Sargent (1972) showed its exceptional
nature. Studies of its colors and resolved stars have often centered on
asking whether there is evidence for any star formation before the
recent burst which dominates its luminosity. The question is whether
I Zw 18 has only recently come into view as a galaxy of stars, or
whether this is simply a very extreme example of a starburst in
a previously faint but gas-rich dwarf system. Whatever its previous
history, at such low metallicities, the abundance pattern is
dominated by only a sparse history of enrichment, making it
potentially a useful analog to chronologically young galaxies
seen at high redshifts.

Table 1 lists some of the most prominent ionic absorption species
seen in the far-UV, identified as to main source (photospheres,
winds, or interstellar absorption). Many have multiple sources
(particularly from either winds or the ISM). The wavelength pattern
shows the blending of several otherwise promising sets of
features which means that only about five lines are both
strong enough and isolated enough to examine in the (low-dispersion)
data available for high-redshift galaxies. The NGC 604 data in
particular suggest, via the wavelength centroid, that P V at 1128 \AA 
is significantly stronger than the adjacent Si IV feature (which
is why the former identification is used here).

Although some velocity confusion exists between M33 absorption and
a foreground high-velocity clouds, some interstellar features
seen in the NGC 604 spectrum appear to arise in the surrounding disk gas. 
Likewise, there are some interstellar lines from the ISM in
I Zw 18 (although most of the prominent ones seen in Fig. 1
arise on our own galaxy).
These spectra provide only upper limits to intrinsic H$_2$ absorption,
at column densities less than $10^{14}-10^{15}$ cm$^{-2}$, in agreement
with the analysis of the same M33 spectra by Bluhm et al. (2003).
The roles of both kinds of ISM absorption are important in comparison
with LBG spectra, where both aperture and column-density effects enter.

The excellent spectral resolution of the FUSE data helps clarify some 
potential issues of line blending for actual hot stellar
populations. For example, they show that the P V $\lambda 1128$ feature, which
for most stellar types is photospheric, is much
stronger than the adjacent Si IV 1122, 1128 pair or Si III $\lambda 1113$.
This contrasts with the situation in the LBGs, where interstellar
absorption lines are much stronger and in some cases overlap with
diagnostic abundance-sensitive lines. These interstellar features,
generally showing strong outflows, are the major difference 
in comparing the local star-forming objects with Lyman-break
galaxies. The column density of foreground gas is an extensive
property, which naturally tends to increase with galaxy scale, making
this one respect in which individual luminous star-forming regions
may not be adequate analogs of Lyman-break galaxies.

\subsection{Lyman-break galaxy spectra}

The FUSE spectra are compared to the group-composite spectra of Lyman-break galaxies (LBGs) discussed
by Shapley et al. (2003), spanning the emitted wavelength range
912-1200 \AA\ . The major variation among spectra of LBGs is in
the net strength of Lyman $\alpha$ emission, which has variously been
attributed to effects of metallicity, geometry, and the importance of global 
winds. As shown in Table 3 of Shapley et al., the sequence
of Lyman $\alpha$ emission is also statistically a ranking by
continuum color (and hence reddening).
The composite spectra include 811 galaxies at
$z \approx 3$, summed in quartiles (roughly 200 galaxies each) 
of Lyman $\alpha$ equivalent width.
The sum not only increases the signal-to-noise ratio of spectral
features in common, but is crucial in averaging out the effects of
absorption in the Lyman $\alpha$ forest. The quality of these spectra varies, 
since there is a rough inverse relation between Lyman $\alpha$ EW and 
continuum luminosity.

I also consider a set of composite spectra grouped into three bins by mid-UV
luminosity. Given the present-epoch increase of metallicity with
luminosity, this subsampling might trace chemical history more purely
than the division on Lyman $\alpha$ strength, which may be driven
by such transient properties as global winds or dust distribution.
However, this division is subject to dust effects, so that one
may fairly wonder whether the UV luminosity scales strongly
enough with star-formation rate, less alone stellar mass, to reflect
any systematic chemical behavior. Given that the UV luminosity
must be high enough for selection as an LBG, we might at least
test for luminosity-linked effects. Nearby galaxies studied with
{\it IUE} do show evidence that the mid-UV luminosity has at
least an envelope, and perhaps a broad correlation, with
emission-line abundances (Calzetti et al. 1994, Keel et al. 2002).
These luminosity composites are derived from the subset of LBGs 
near the peak of
the color-selection function with redshift, $2.9 < z < 3.1$, and
will unavoidably reflect spectroscopic biases (most notably the
easier measurement of redshifts for emission-line objects at the 
faintest levels, discussed by Shapley et al. 2003). Absolute
magnitudes are evaluated in the AB$_\nu$ system near 1700 \AA\ ,
using a cosmology with H$_0 = 71$ km s$^{-1}$ Mpc$^{-1}$,
$\Omega_m=0.3$, $\Omega_\Lambda = 0.7$.
The faintest bin, 
spanning from $M_{UV}$=-20.55 to -19.52, includes
90 galaxies. The intermediate-luminosity bin samples from 
$M_{UV}$=-21.02 to -20.56
with 93 galaxies, and the brightest group includes 92 galaxies from
$M_{UV}$=-22.58 to -21.03. Average $M_{UV}$ values are -20.25, -20.79, 
and -21.44 for these subgroups. As shown in Fig. 3, some strong
absorption features do exhibit systematic trends with UV luminosity.
The O VI doublet and both N II and N III (the latter possibly blended
with interstellar O I) all strengthen for more UV-luminous galaxies,
results which suggest that there may indeed be composition changes
along this sequence. The signal-to-noise ratios of the composite
spectra are naturally rather different; there are only rather loose
upper limits for the bluest lines ($\lambda \lambda 955, 991$)
for the fainter groups.

The effective resolution of these spectra is limited both by the
uncertain placement of the galaxies within fairly wide slits and by
uncertainty in the individual redshifts, since global winds 
and radiative-transfer effects can make redshifts measured from
Lyman $\alpha$ or ISM lines alone differ from the actual systemic value.
I estimate the resolution of each spectrum through Gaussian fits
to absorption features dominated by interstellar components, particularly
Fe II and Si II. Most such lines are in the mid-UV from 1260-1670 \AA\ ;
several have at least weak P Cygni wind-emission components. The
overall average FWHM of these lines is 3.3 \AA\ , with a trend
for narrower profiles in the stronger-emission composites. 
However, some of these lines are weaker in the objects with strong
Lyman $\alpha$ emission, so these measurements are correspondingly
less certain. The effective line width may be dominated in some galaxies by the
width of absorption lines from outflowing gas, which reaches a detected
span of 1300 km s$^{-1}$ in at least MS1512-cB58 (Pettini et al. 2002).
Such widths would apply to interstellar absorption but not to stellar features.


\section{Stellar Abundances}

In contrast to
the nebular emission lines which are prominent in the optical regime,
the continuum from massive stars and both emission and absorption lines from
their winds dominate the UV, and particularly far-UV, spectral region 
of star-forming galaxies.
Reliance on absorption features makes abundance analyses observationally 
challenging, since emission lines may place much looser demands on the
signal-to-noise ratio in the continuum.
However, the results are astrophysically easier to interpret,
since we are guaranteed to measure the composition of the massive
stars themselves at the epoch observed, rather than some mixture of pristine
and enriched interstellar material. The strong stellar-wind lines have particular
promise as metallicity indicators, since abundances enter both into
the fraction of the wind column density from each element and 
into the overall mass-loss rate in radiatively-driven winds.
This notion has been applied by Pettini et al. (2000) and Leitherer
et al. (2001), the latter using a synthetic library of model spectra.
Mehlert et al. (2002) used the C IV $\lambda 1550$ 
doublet to derive rough abundances for a sample of LBGs. They found 
carbon abundances evolving in the mean from 0.16 to 0.42 solar over the
range $z=3.2-2.3$, but use of the C IV lines alone, at low resolution, is
likely to be compromised by the role of absorption from
outflowing material, and be affected by
the star-forming history and geometry of the galaxies, through possible
variations in ionization balance and relative amounts of emission and 
absorption. 
Use of the far-UV range can expand this approach to a wider range of
elements and ionization states. The
grounds for comparison are shown in Figs. 2 and 3, which show the locations
of wind, photospheric, and interstellar lines on the two sets of
composite LBG spectra, in comparison to
a similarly smoothed version of the NGC 604 FUSE spectrum.
In particular, the LBG composites show stronger interstellar atomic 
absorption than any of the local FUSE targets, which for some features
limits the accuracy of derived absorption equivalent widths for the
stellar features.

Keel et al. (2004) used the FUSE results to propose an empirical
relation between the equivalent widths of far-UV photospheric and wind
lines and the overall metallicity, parameterized through [O/H]
as measured from the optical emission lines. These equivalent widths
are with respect to local pseudocontinuum bands selected from the
high-resolution FUSE spectra to be free of discrete absorption features,
bands which were matched as nearly as possible in measuring the
LBG composite spectra.
Not all strong stellar-wind lines in the far-UV
are amenable to such analyses; for example, the components of the O VI doublet
are usually thoroughly blended, with each other and with interstellar absorption of C II and
Lyman $\beta$. Useful relations
exist for C II, C III, C IV, N II, N III, and P V. Among these,
P V has a substantial photospheric contribution for many 
stellar types and has the most nearly linear EW-(O/H) behavior.
As a group, all these relations should yield the same results
only for galaxies with a long history of stellar nucleosynthesis and
recycling, since various elements should enter the ISM and be incorporated
into stars at quite different rates. In fact, one can use changes 
in the abundances derived from various species to seek signs of
this history in high-redshift galaxies. Carbon and phosphorus are
primary elements (although the yield may depend on
stellar abundances, from the work of Maeder 1992), 
while the situation for nitrogen remains ambiguous
at low metallicity.

Within the LBG composite spectra, the major changes with
Lyman $\alpha$ strength are in the P V $\lambda 1122$ and
N II $\lambda 1084$ lines, both of which weaken with stronger Ly $\alpha$.
Tables 2 and 3 summarize the line properties of the LBG spectra,
grouped by Ly $\alpha$ strength and by mid-UV luminosity.
At this spectral resolution, absorption and emission components are
seldom well separated for wind features, so net equivalent
widths are listed. These should be robust to the resolution of the data as long
as there are not issues of other spectral features contaminating the
``continuum" windows. As in deriving the empirical EW-metallicity
relations, I have been guided by the FUSE spectra in selecting lines
for which such blending issues are least important.

As an extreme example of potential absorption blends, 
the Si II $\lambda 1526$ line is nearly as strong as
C IV wind absorption in most of the LBG composite spectra. Using 
large-aperture data for objects in the local sample 
from the IUE archive (as collected by Keel et al. 2004)
to sample apertures comparable
to the FUSE data, the ratio
of equivalent widths between Si II and C IV absorption components
falls in the range 0.38-0.50. In one LBG, MS 1512-cB58 at
$z=2.728$, gravitational lensing has allowed unusually detailed
spectra to be obtained using Keck (Pettini et al. 2002) and the VLT 
(Savaglio, Panagia, \& Padovani 2002), in an excellent example
of Zwicky's (1937) ``gravitational telescope".
Far-UV lines in cB58 can suffer from confusion with the absorption in the
Lyman $\alpha$ forest at this redshift, although upper limits
can be set for emitted wavelengths longward of 1073 \AA\  for those
lines which do not overlap an obvious lower-redshift Lyman $\alpha$ feature.
However, these data are very informative about absorption lines
longward of Lyman $\alpha$. The Si II/C IV ratio of 0.48 (including the
diluting effects of C IV emission, to mimic lower-resolution
data) lies at the high end of what we see in local objects.

The N II $\lambda 1083$ region in cB58 is more complex. Fairly
broad absorption is present roughly centered on the expected
wavelength of the (blended) N II triplet, with a strong narrow
feature centered at $\lambda_0 = 1080.0$ \AA. This might be
Fe II $\lambda 1081.87$ blueshifted by $\approx 500$ km s$^{-1}$;
if not, it would be the strongest observed feature in the Lyman $\alpha$
forest and this Fe II would be undetectably weak. This narrow,
potentially interstellar component contributes a fraction 0.25 of the equivalent
width of the whole complex, and would be thoroughly blended with the
N II feature in our composite spectra.

The equivalent widths in Tables 2 and 3 have been combined with the empirical
relations from Keel et al. (2004) to estimate abundances for
each of these species. For comparison with optical emission-line 
results, these relations use [O/H] as the independent variable.
While convenient, it does leave the meaning of these abundances
slightly opaque (being the abundance of that species typically
encountered at the given oxygen abundance in present-day galaxies).
Comparing the multiple ions available for C and N is useful, since the 
ionization state of winds will generally change with stellar abundances; the
mean is listed for each as well as individual values. 

Effects of
unresolved interstellar absorption lines will all increase the
derived abundances, but the amplitude of the overestimate depends
strongly on the mean velocity offset and width of this
absorption component. A narrow component near the maximum
stellar-wind absorption may have little effect since there is
little light to absorb at that wavelength, while broad and
blueshifted outflow profiles can demonstrably contribute a
substantial part of the overall equivalent width. For resonance
lines, this issue means that these derived abundances must strictly
be considered as upper limits. This has been reinforced by
the analysis of Crowther et al. (2006), in showing the
importance of accounting for interstellar absorption in
the C IV transitions before deriving abundances to avoid a systematic
overestimate. Once again, the scale of the system and its ISM can
make a significant difference between spectroscopic properties of
individual H II regions and entire galaxies (as seen for the LBG
samples). In this comparison, the data for NGC 7714 and I Zw 18 
encompass almost all their UV flux, while NGC 604 contributes
nearly 20\% of the entire far-UV flux from M33 (from
{\it Voyager} UVS data, in Keel 1998). Still, the available
data point to interstellar absorption in LBGs which
is stronger and broader than in the most nearly comparable
nearby systems, and often blueshifted compared to photospheric absorption
or nebular emission lines, 

The implied abundance levels are given in Tables 4 and
5, again for the Ly $\alpha$ and luminosity groupings. 
The C and N abundances
are roughly half solar, with no evidence of a trend with Lyman $\alpha$
strength. In this respect, the luminosity groupings may be
more useful, since the numbers show a significant increase in
the C lines with luminosity. There is an increase in the
upper limit for N as well, with the limit set by the fact
that N II seems to have a floor for [O/H] below 0.48 as
well as the limited sensitivity for the fainter groupings.
 
This is a milder C/O deficit than inferred from
mid-UV emission lines by Shapley et al. (2003). 
In contrast to the C and N results,
the P abundance is too low to measure (less than
0.1 solar) except possibly in the reddest and weakest-emission 
LBGs (group 1) and the intermediate-luminosity composite. 
This lack of absorption is observationally robust, possibly even
exacerbated by the role of interstellar absorption,
and suggests that these objects are in an early stage of nucleosynthetic
history as far as this element is concerned. Phosphorus is a product of
the reaction network in red-giant interiors, essentially from the
combination of $\alpha$- and $s$-processes,
with some contribution from supernovae (dominated by type II
explosions), as summarized by Trimble (1991). The high oxygen
masses expected for SN II remnants might lead to O/H measures
that are higher than would be expected from scaling other elements as well,
during the interval from the onset of SN II enrichment until the
evolution of solar mass stars has time to approach the enrichment that
our vicinity leads us to call normal.

Other evidence has suggested that the abundances in LBGs are low for
luminous galaxies, but not extremely low. In addition to the C IV
wind data of Mehlert et al. (2002), the traditional optical
bright-line technique has been applied via near-infrared spectroscopy.
Pettini et al. (2001) showed that the gas-phase O/H
ratio must typically lie above 0.1, from the [O III]/H$\beta$ ratios
alone. For a few selected objects with more extensive spectral coverage,
Kobulnicky \& Koo (2000) had found O/H = 0.25-0.95 solar, fitting with the
more limited results of Teplitz et al. (2000) from a combination
of spectroscopy and fortuitously placed filter imaging. These
results for LBGs stand in marked contrast to the low abundances found
for damped Lyman $\alpha$ absorption systems, indicating a marked
divergence in the history of these disparate environments. They
also contrast with the fainter narrow Lyman $\alpha $ emitters for
which colors and limited spectroscopic data suggest still lower
abundances, O/H below 1/3 solar and conceivably as low as in I Zw 18
(Keel et al. 2002).

\section{Molecular Hydrogen Absorption}

The far-UV range offers a unique way to probe molecular gas in these
young systems, sensitive even to cold H$_2$ and biased only by the
requirement that sufficient UV continuum penetrates the molecular medium
to contribute to the emerging UV flux. This bias may account for the
low limits to H$_2$ column density derived for the M33 H II regions
by Bluhm et al. (2003) and for I Zw 18 by Aloisi et al. (2003). In
each case, analysis of FUSE data indicates N(H$_2$) $< 10^{17}$ cm$^{-2}$
in all cases, and $< 10^{14}$ for I Zw 18, in which the wavelength
overlap of H$_2$ multiplets from the target galaxy with other features
in Milky Way material is less serious. Hoopes et al. (2001)
suggest that similar column densities apply to more luminous
starburst systems, and implicate the UV-transparency bias. This bias
should be minimized at low abundances, where the optical and UV extinctions are 
smaller for a given gas column density (Testor \& Lortet 1987;
Quillen \& Yukita 2001). The persistence, or absence, of such a bias
could trace the development of the gas-to-dust ratio as elements needed 
for grain formation are progressively enriched in the interstellar medium.

The group composite spectra of Lyman-break galaxies can be examined for possible 
signatures of H$_2$ absorption. At their resolutions,
even the strongest individual H$_2$ features are blended, so I
consider what combinations of H$_2$-free template spectra
(from the low-redshift objects and theoretical predictions) 
fit the LBG spectra most closely
after folding in H$_2$ absorption spectra for various column densities.
Single column density values may not reflect the average across the
galaxy closely, given the complex weighting of both the starlight and
has distributions (as discussed by Bluhm et al. 2003) and the
possibility that actively star-forming galaxies might have multiple
velocity systems which strengthen the overall absorption.
Still, even crude values are
useful given the orders-of-magnitude disparity between typical Milky Way sight 
lines and values derived for bright star-forming regions in nearby galaxies.
The ``local" templates come from the spectra of NGC 604 and I Zw 18
after removal of galactic H$_2$ absorption. The removal was done by
fitting and subtracting Gaussian profiles (for weak isolated lines)
or Voigt profiles (with the same parameters for members of each strong multiplet)
at the wavelengths of all individually detected H$_2$ lines from the
compilation by McCandliss (2003), except those lines confused with 
expected stellar-wind features. The ratio of spectra before and after
this process gives the spectrum of foreground Galactic absorption
sampled along these lines of sight, which we can use as empirical templates for
H$_2$ absorption. I also consider purely theoretical templates, using the
absorption spectra from McCandliss (2003) for intrinsic velocity width
$b$= 10 and 20 km s$^{-1}$. The H$_2$OOLS IDL routines generate
such templates for various sets of level populations, showing 
that the levels $J=0-2$ are well populated in the galactic foreground
and would likely be so in other galaxies as well. Populations with
$J>2$, although much less populated than the lower levels,
are generally seen to be populated well above the expected
thermal levels based on the lower states, which is attributable to radiative
excitation from the ambient starlight.

Some features in the composite LBG spectra do fall
near H$_2$ bands, although the opportunities for confusion with relatively
weak atomic lines are many (Fig. 4). The matches change with assumed H$_2$
rotational temperature, since adjacent bands shift slightly to the
red for transitions arising at higher $J$. This effect amounts to
typically 6 \AA\  between $J=0,3$, and the higher $J$ absorption
has a more complex profile (which will be broader even at the relatively
coarse resolution of the composite spectra).
This structure is one of the reasons that absorption from these 
higher-lying states is stronger for a given column density.
The strongest constraints come from wavelengths with no observed absorption,
where H$_2$ absorption models predict features comparably deep to the
ones observed. The most prominent of these are near 1002, 1050, and 1078 
\AA\ . On the other hand, there are features matching some of the strongest
H$_2$ bands in otherwise clean spectral regions near 986, 1011, 1064, and 1096 
\AA\ . Weak interstellar lines, not seen in NGC 604, do occur near these
features as follows: O I 988, S III 1012, Si IV (wind) 1062, Fe II 1063,
and Fe II 1096. The band structure leads to the features near 987 and 1012
becoming relatively stronger for warmer H$_2$. This leaves two bands
near 1050 and 1092 \AA as the 
most sensitive regions for seeking H$_2$ absorption. The highest
signal-to-noise here occurs in the group 1 Ly $\alpha$ composite,
with equivalent-width limits for these features of 0.27 and 0.20
\AA respectively. 

For the group 1 composite spectrum, comprising the LBGs with net
absorption at Lyman $\alpha$, regions of the spectrum not
affected by strong atomic lines can be matched with a synthetic
spectrum for N(H$_2$) = $1.7 \times 10^{18}$ cm$^{-2}$ if the
molecules are mostly in $J=3$; this fit does not violate any
spectral regions with high local S/N. This $J$ comes from the mean wavelengths
of spectral dips near 1098 and 1114 \AA\ , as well as the strength
of the 987 and 1012-\AA\  features. These data rule out comparable
contributions from lower states at more stringent equivalent-width
limits for
decreasing $J$. However, the limits on N(H$_2$) in the lower
states are comparable, since the smoothed absorption per unit
column density is weaker for these transitions.

For the group-1 composite, one could view this as evidence for
a substantial amount of warm molecular absorption, or as an upper limit
including lower states of $\approx 7 \times 10^{18}$ cm$^2$.
This is a very modest column density by the standards of Milky Way
sight lines; the FUSE spectra of I Zw 18 and the M33 H II star-forming
regions both show foreground absorption at  N(H$_2$) = $\approx
5 \times 10^{18}$ cm $^{-2}$ at galactic latitudes $b = 44^\circ$
and $-31^\circ$ respectively. In both of these objects, low limits
to internal H$_2$ absorption have been set from FUSE data
(Aloisi et al. 2003, Bluhm et al. 2003, Keel et al. 2004).
The cleanest measurement is $< 10^{14}$ cm$^{-2}$ for I Zw 18,
at a large enough redshift to reduce contamination by Milky Way
absorption. Limits for the M33 regions are $< 10^{17}$ cm$^{-2}$
in all cases. All these values must be affected by the UV bias;
the mixing of molecular gas with dust means that stars behind
large amounts of H$_2$ will not contribute to the overall
UV flux.

The other groups of LBGs have slightly lower limits on H$_2$
absorption than the possible detection in the group 1 objects.
This result indicates that the ratio of UV extinction to molecular-gas
column density (that is, the dust-to-gas ratio)
was not much smaller in the Lyman-break galaxies than
in local star-forming systems, so that stars behind substantial
amounts of H$_2$ are likewise obscured and vanish from the UV
spectrum as well. This in turn suggests that any large
amounts of H$_2$ in Lyman-break galaxies are mixed with dust
(and by implication, the whole range of metals) rather than
being primordial and unenriched, because it is just the latter
case which would show up most effectively 
via strong absorption in the Lyman and Werner bands.

The Lyman $\alpha$ forest will limit our ability to derive more precise
H$_2$ column densities, since in any individual object at high redshift,
there will be confusion between its absorption and random H I features.
We may be limited to such rough``typical" class properties.

\section{Conclusions}

Far-UV spectra of nearby star-forming galaxies are compared
to composite spectra of Lyman-break galaxies at $z \approx 3$, with 
special attention to abundance-sensitive features. Features of C and N
suggest moderately subsolar abundances, at the level associated with
O/H roughly half solar in present-day galaxies. There is some evidence that
the C abundances correlate with mid-UV luminosity, and N may do
so subject to detection limitations that introduce upper limits.
Phosphorus is underabundant
by an order of magnitude in most subsamples of LBGs, perhaps reflecting
a dominant origin in red giants rather than SN II. This fits with
additional evidence from emission-line measurements, that the ionized
gas is of low but not very low abundances, and suggests in itself that the
star-forming histories of LBGs have been brief but eventful.

The composite spectra also allow limits on the column density
of H$_2$ against the UV starlight in typical LBGs, which are comparable
to what we see looking out of the Milky Way at modest latitudes
$l \approx 30^\circ$. While some bias toward unextinguished stars
must exist, this suggests that there are not huge reservoirs of cold and
unenriched gas surrounding LBGs.

\acknowledgments
I am particularly grateful to Alice Shapley and Chuck Steidel for
providing the composite LBG spectra in digital form, and for
extensive and challenging discussions.
This work was supported in part by NASA FUSE grant NAG5-8959,
with publication support kindly enhanced by George Sonneborn at the
NASA FUSE office.
I acknowledge the community service rendered by Steven McCandliss
in making his H$_2$OOLS software and data compilation publicly accessible. 
Alessandra Aloisi provided the summed I Zw 18 spectra, and B.-G. Andersson
was helpful in planning and reducing the FUSE observations.

\begin{figure}
\epsscale{0.5}
\includegraphics[scale=0.70,angle=90]{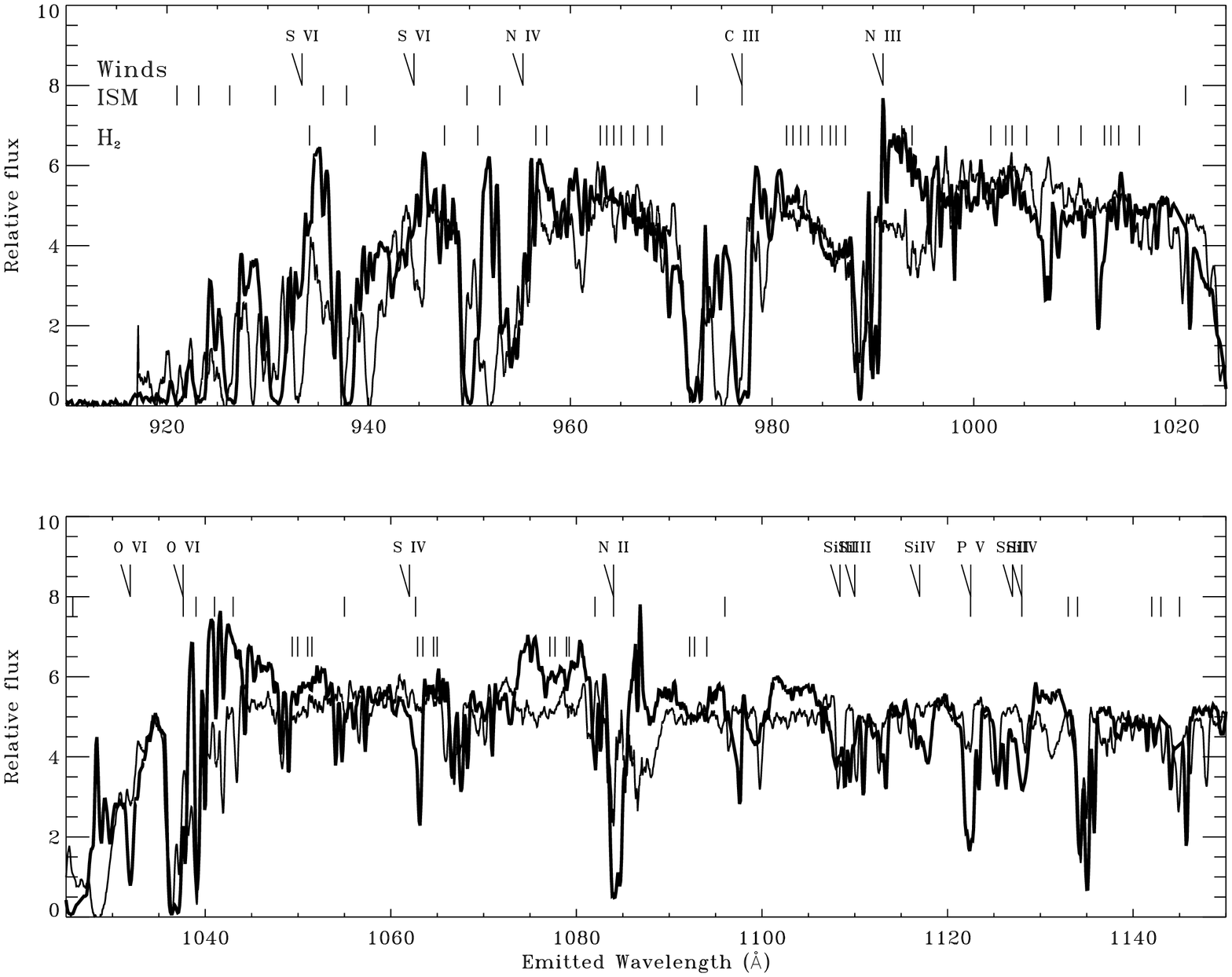}
\caption
{A field guide to far-ultraviolet galaxy spectra. The FUSE spectra of NGC 604
(thick line) and I Zw 18 (thin line) are shown, with Galactic H$_2$ absorption 
lines fitted and removed, and
smoothed by 0.3 \AA\  for graphic simplicity, with various
line wavelengths illustrated. Wavelengths of the strongest H$_2$ transitions
are shown by the lowest set of marks; the line list is the set that was
strong enough for Milky Way absorption to be detected against NGC 604.
The strongest atomic absorption features from the interstellar medium
are marked in the middle row. Lines with significant stellar-wind
contributions are marked at the top with schematic blueward wings on
each symbol, and ions identified. Lyman $\beta$ lies at the break between 
panels. The relative flux is in F$_\lambda$, and represents the observed
flux in erg cm$^-2$ s$^{-1}$ \AA\ $^{-1}$ scaled upward by $10^{13}$
for NGC 604 and $2 \times 10^{14}$ for I Zw 18. Differences between the two
spectra include stronger stellar-wind features, in both emission and
absorption portions of the P Cygni profiles, in NGC 604, stronger
absorption in P V from NGC 604 for which the dominant contributor is
stellar photospheres, and a few apparently blueshifted interstellar
features in I Zw 18 which are foreground Milky Way gas. The wavelength 
scale is the emitted frame for each object.
}
\end{figure}

\begin{figure}
\epsscale{0.5}
\includegraphics[scale=0.70,angle=90]{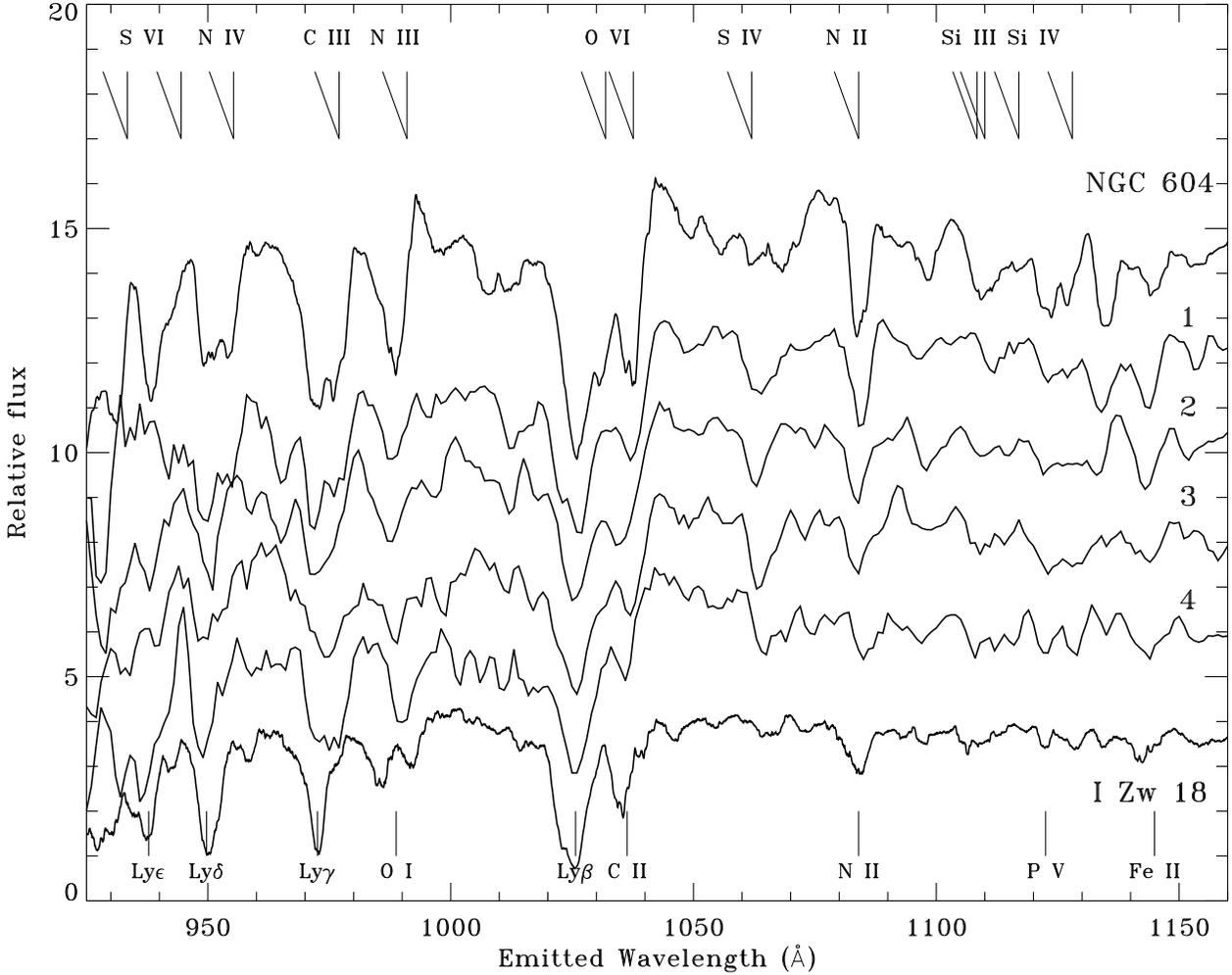}
\caption
{Comparison of the composite LBG spectra with the FUSE data on NGC 604 and
I Zw 18,
after foreground atomic and H$_2$ absorption lines have been removed.
The LBG groups are numbered as in Shapley et al. (2003) and Table 2,
with group 1 having net Lyman $\alpha$ absorption, and 2-4 with
stronger net emission in that order. All the spectra have been boxcar-smoothed 
to a common resolution $\approx 4.1$ \AA\ 
for display, a value set by convolving the LBG composites with their typical
resolution to enhance the visibility of weak lines. The flux zero point is 
for the lowest spectrum; successively
higher ones are offset by integer values units. Important interstellar 
and photospheric features are marked at the bottom; wind lines are shown
across the top. The symbols accompanying the wind designations are intended
to suggest a blueshifted absorption trough, but some of these lines
have distinct redshifted emission components as well (often blended
with components of additional lines, as in the O VI doublet).
}
\end{figure}

\begin{figure}
\epsscale{0.5}
\includegraphics[scale=0.70,angle=90]{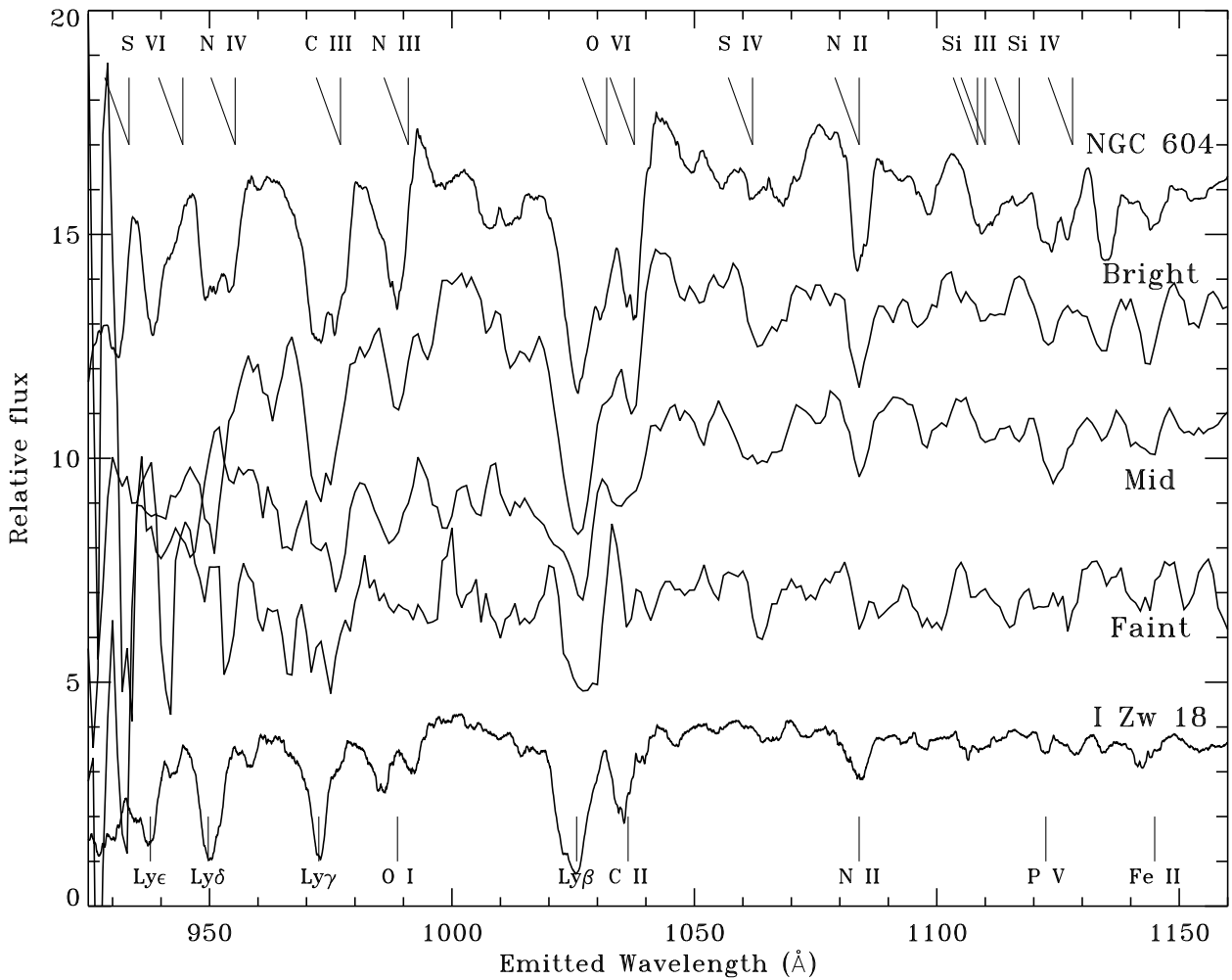}
\caption
{Comparison of the composite LBG spectra, now grouped in three
bins of mid-UV luminosity, with the FUSE data on NGC 604 and I Zw 18,
displayed as in Fig. 2.
Several strong lines show systematic trends with UV luminosity.
For example, the O VI doublet,
N II $\lambda 1083$, and N III $\lambda 991$ 
all strengthen for brighter galaxies,. 
Brighter galaxies are
also redder in continuum slope shortward of about 1000 \AA .}
\end{figure}

\begin{figure}
\epsscale{0.5}
\includegraphics[scale=0.70,angle=90]{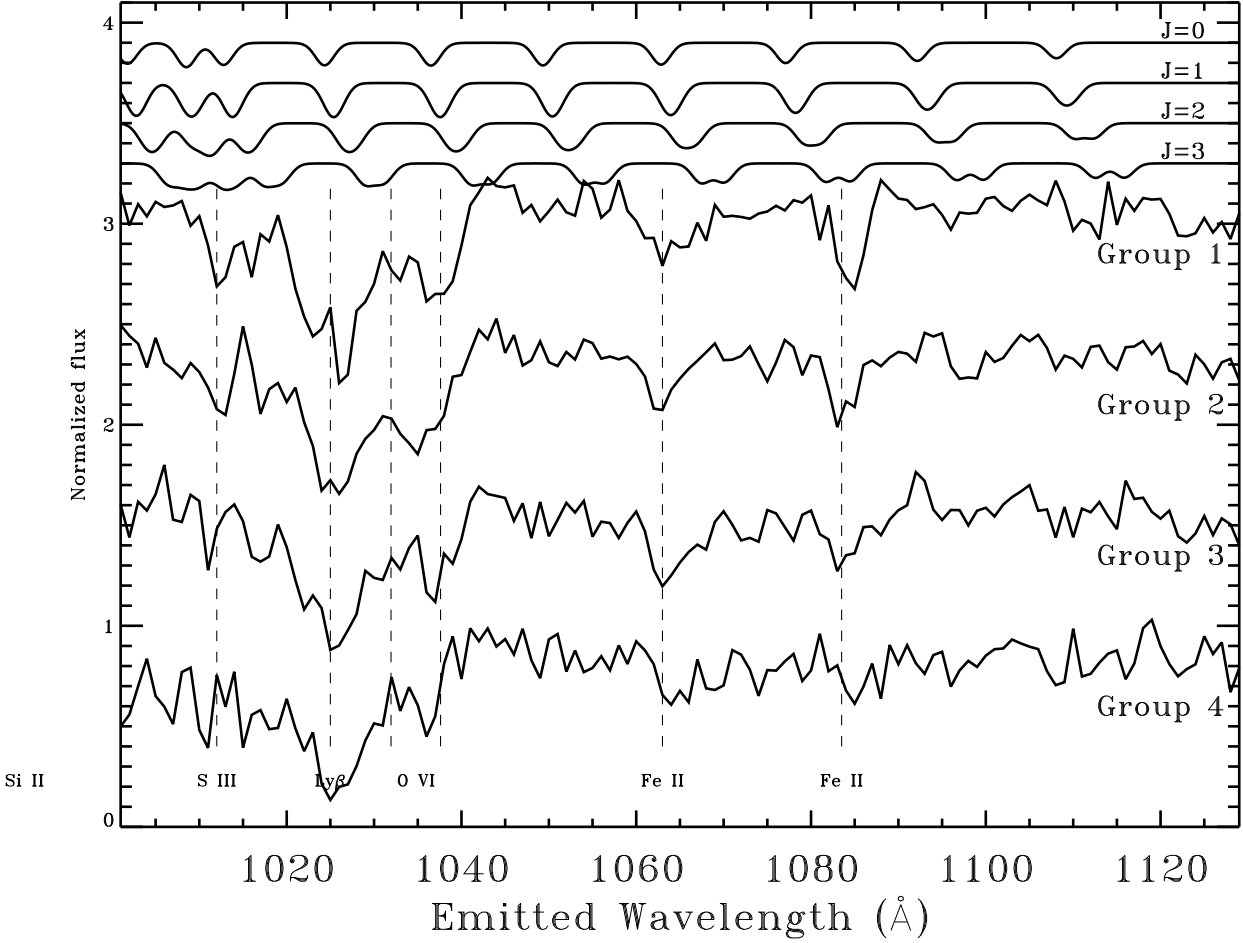}
\caption
{Composite LBG spectra, with the continuum shape normalized to
indicate absorption depths, compared to theoretical absorption spectra,
as calculated with S. MacCandliss' H$_2$OOLS routines. Each $J$
value is shown separately, for a column density $1.7 \times 10^{18}$
cm$^{-2}$. This value is the maximum that any $J$ population could
have against the group 1 spectrum without violating the data, although
having populations weighting strongly to $J=3$ is physically
implausible. Potentially
confusing ionic absorption wavelengths are labelled at the bottom.}
\end{figure}

\clearpage
\begin{table*}
\begin{center}
\begin{tabular}{cclccl}
\tableline
\tableline
$\lambda$ (\AA ) & Species & Source &  $\lambda$ (\AA ) & Species & Source \cr
 933.4  & S VI & winds & 1037.6  & O VI & winds+ISM \cr
 944.5  & S VI & winds & 1039.2  & O I & ISM \cr
 949.7  & Ly $\delta$ & phot+winds+ISM & 1048.2 & Ar I & ISM \cr
 953.9  & N I & ISM & 1062.7  & S IV & winds \cr
 955.3  & N IV & winds & 1063.2  & Fe II & ISM \cr
 971.7  & O I & ISM & 1063.4  & S I & ISM \cr
 972.5  & Ly $\gamma$ & phot+winds+ISM &1066.7 & Ar I & ISM \cr
 976.4  & O I & ISM & 1081.9  & Fe II & ISM \cr
 977.0  & C III & phot+winds+ISM & 1084.6 & N II$^3$ & winds/ISM \cr
 988.7  & O I$^3$ & ISM & 1096.9  & Fe II & ISM \cr
 989.8  & N III$^*$ & winds+ISM & 1108.4 &  Si III & winds \cr
1012.5  & S III & winds/ISM & 1110.0  & Si III & winds \cr
1020.7  & Si II & ISM &  1113.2  & Si III & winds \cr
1025.8  & O I$^3$ & ISM & 1122.5  & Si IV & winds/photospheres \cr
1025.7  & Ly $\beta$ & phot+winds+ISM & 1128.0 & P V & photospheres/winds \cr
1031.9  & O VI & winds/ISM & 1128.3  & Si IV & winds/photospheres \cr
1036.3  & C II$^*$ & ISM & 1134.4  & N I$^3$ & ISM \cr

\tableline
\tablecomments{Wavelengths are as emitted. The dominant source of each
line is given as photospheric, winds, or ISM, following the
attributions in Leitherer et al. (2002), Walborn et al. (2002),
and Danforth et al. (2002). A superscript 3 denotes that
the listed wavelength is for the central member of a triplet,
and an asterisk indicates absorption from an excited state.}
\end{tabular}
\tablenum{1}
\caption{
Prominent Far-UV Features in Galaxy Spectra\label{tbl1}}
\end{center}
\end{table*}

\begin{table*}
\begin{center}
\begin{tabular}{lcccc}
\tableline
\tableline

Group: & Ly $\alpha$-1  & Ly $\alpha$-2 & Ly $\alpha$-3 & Ly $\alpha$-4 \\
Ly $\alpha$           & 14.8 & 4.2 & -8.5 & -46.6 \\ 
N IV $\lambda 955$    &  3.3 & 1.9 & 1.9 & 1.9  \\
N III $\lambda 991$   &  1.1 & 1.5 & 1.0 & 1.8  \\
N II $\lambda 1083$   &  1.3 & 1.0 & 0.9 & 0.5 \\
C IV $\lambda 1549$   & 3.45 & 2.55 & 2.83 & 2.30 \\
C III $\lambda 977$   & 3.6 & 2.4 & 2.5 & 3.6  \\
P V $\lambda 1118/25$ & 0.6 & 0.2: & 1.3 & $<0.5$  \\
\tableline
\tablecomments{Equivalent widths in \AA\  in the emitted frame; positive
values represent absorption.}
\end{tabular}
\tablenum{1}
\caption{
Far-UV Absorption Features in Emission-Grouped Composite LBG Spectra\label{tbl1}}
\end{center}
\end{table*}

\begin{table*}
\begin{center}
\begin{tabular}{lccc}
\tableline
\tableline

Group: & Bright & Intermediate & Faint \\
Ly $\alpha$           & 1.34 & -2.72 & -26.3    \\ 
N IV $\lambda 955$    & 3.08 & ...   & $<0.6$   \\
N III $\lambda 991$   & 1.19 & 1.0   & $<0.6$   \\
N II $\lambda 1083$   & 0.98 & 1.17  & 0.56     \\
C IV $\lambda 1549$   & 3.5  & 4.3   & 2.4      \\
C III $\lambda 977$   & 4.0  & 1.9   & 0.9      \\
P V $\lambda 1118/25$ & 0.40 & 1.06  & 0.45     \\
\tableline
\tablecomments{Equivalent widths in \AA\  in the emitted frame; positive
values represent absorption.}
\end{tabular}
\tablenum{2}
\caption{
Far-UV Absorption Features in Luminosity-Grouped Composite LBG Spectra\label{tbl1}}
\end{center}
\end{table*}

\begin{table*}
\begin{center}
\begin{tabular}{lcccc}
\tableline
\tableline

Group: & Ly $\alpha$-1  & Ly $\alpha$-2 & Ly $\alpha$-3 & Ly $\alpha$-4 \\
C IV            & 0.13 &    0.09    & 0.11   &   0.07   \\
P V             & 0.39 &    $<0.05$ & 0.10   &   $<0.10$  \\
N IV            & 0.80 &    0.62    & 0.61   &   0.62  \\
C III           & 0.9  &    0.77    & 0.78   &   0.9  \\
N III           & 0.42 &    0.56    & 0.46   &   0.60  \\
N II            & 0.57 &    0.50    & $<0.5$ &   $<0.5$  \\
C III + IV      & 0.52 &    0.43    & 0.45   &   0.48  \\
N II + III + IV & 0.60 &    0.56    & 0.52   &   0.57  \\
\tableline
\tablecomments{Abundances are in solar units, with a local calibration
referred to nebular [O/H] measurements. Multiple ions are combined
by averaging in the bottom lines.}
\end{tabular}
\tablenum{3}
\caption{
Derived Effective [O/H] for Emission-Grouped Composite LBG Spectra\label{tbl1}}
\end{center}
\end{table*}

\begin{table*}
\begin{center}
\begin{tabular}{lcccc}
\tableline
\tableline

Group: & Bright  & Intermediate & Faint  \\
C IV            & 0.14 & 0.21 & 0.07 \\
P V             & 0.18 & 0.77 & 0.20  \\
N IV            & 0.79 & $<0.35$ & $<0.35$ \\
C III           & 1.2  & 0.71 & 0.42 \\
N III           & 0.52 & 0.47 & 0.40  \\
N II            & $<0.48$ & $<0.48$ & $<0.48$ \\
C III + IV      & 0.45 & 0.31 & 0.18  \\
N II + III + IV & $<0.60$ & $<0.33$ & $<0.38$ \\
\tableline
\tablecomments{Abundances are in solar units, with a local calibration
referred to nebular [O/H] measurements. Multiple ions are combined
by averaging in the bottom lines.}
\end{tabular}
\tablenum{4}
\caption{
Derived Effective [O/H] for Luminosity-Grouped Composite LBG Spectra\label{tbl1}}
\end{center}
\end{table*}

\end{document}